\documentstyle[epsfig]{jaa}

\DeclareMathAlphabet{\mathsc}{OT1}{cmr}{m}{sc}
\def\testbx{bx}%
\DeclareRobustCommand{\ion}[2]{%
\relax\ifmmode
\ifx\testbx\f@series
{\mathbf{#1\,\mathsc{#2}}}\else
{\mathrm{#1\,\mathsc{#2}}}\fi
\else\textup{#1\,{\mdseries\textsc{#2}}}%
\fi}
\def\ch{\footnotesize}
\def\HI{\ion{H}{i}~}
\def\km{km~s$^{-1}$~}
\def\deg{\hbox{$^\circ$}~}
\def\aa{Astron. Astrophys.}
\def\ApJ{Astrophys. J.}
\def\MNRAS{Mon. Not. R. Astron. Soc.}
\def\AJ{Astron. J.}
\def\ARAA{Ann. Rev. Astron. Astrophys.}

\begin{document}
\title[Eridanus group : \HI content]
{The \HI content of the Eridanus group of galaxies}
\author[Omar \& Dwarakanath] 
{A. Omar\thanks{Present address : ARIES, Manora peak, Nainital, 263 129, 
Uttaranchal, India}
\thanks{e-mail: aomar@upso.ernet.in}
\& K.S. Dwarakanath
\thanks{e-mail: dwaraka@rri.res.in}\\
Raman Research Institute, Sadashivanagar, Bangalore 560 080, India \\}

\pubyear{xxxx}
\volume{xx}
\date{Received xxx; accepted xxx}
\maketitle
\label{firstpage}
\begin{abstract}

The \HI content of galaxies in the Eridanus group is studied using the GMRT
observations and the HIPASS data. A significant \HI deficiency up to a factor of
$2-3$ is observed in galaxies in the high galaxy density regions. The \HI
deficiency in galaxies is observed to be directly correlated with the local
projected galaxy density, and inversely correlated with the line-of-sight
radial velocity. Furthermore, galaxies with larger optical diameters are
predominantly in the lower galaxy density regions. It is suggested that the \HI
deficiency in Eridanus is  due to tidal interactions. In some galaxies,
evidences of tidal interactions are seen. An important implication is  that
significant evolution of galaxies can take place in the group environment. In
the hierarchical way of formation of clusters via mergers of groups, a fraction
of the observed \HI deficiency in clusters could have originated in groups.
 The co-existence of S0's and severely \HI deficient galaxies in
the Eridanus group suggests that galaxy harassment is likely to be an effective
mechanism for transforming spirals to S0's.

\end{abstract} 

\section{Introduction}

Spiral galaxies in the cores of clusters are known to be  \HI deficient
compared to their field counterparts (Davies \& Lewis 1973, Giovanelli \&
Haynes 1985, Cayatte et al. 1990, Bravo-Alfaro et al. 2000, Solanes et al.
2001). Several gas-removal mechanisms have been proposed to explain the \HI
deficiency in cluster galaxies. There are convincing results from both the
simulations and the observations that  ram-pressure stripping (cf. Gunn \& Gott
1972) is effective in galaxies which have crossed the high intra cluster medium
(ICM) density region near the core of the cluster (Vollmer et al. 2001, van
Gorkom 2003). ``Galaxy harassment'' can also affect  outer regions of the disk
as a result of repetitive fast encounters of galaxies in clusters (Moore et al.
1998). There can be other scenarios where galaxies can become gas deficient,
e.g., thermal evaporation and viscous stripping (Cowie \& Songaila 1977, Nulsen
1982, Sarazin 1988), and ``galaxy starvation'' where hot gas in the halos of
galaxies is stripped. It is believed that halos contain a reservoir of hot gas,
which sustains star formation in galaxies over their present ages (Larson et
al. 1980). 

Often one or more processes have been shown to be working in individual cases.
There are no strong arguments for any of these processes to be globally
effective in clusters. Cayatte et al (1990) showed that the \HI deficiency in
Virgo galaxies can be understood by a combination of ram-pressure stripping and
transport processes. However, there are several inconsistencies. Magri et al.
(1988) showed that no single gas-removal process can be justified consistently
in any cluster. Further uncertainties arise since some of the parameters
driving these mechanisms are not known well, e.g., thermal conductivity of the
ICM, amount of hot gas in halos etc. It is also not clear that all severely \HI
deficient galaxies have crossed the core as required for ram-pressure to be
effective. Contrary to what is expected from ram-pressure stripping,  the low
mass spirals and dwarfs are indistinguishable from the massive spirals in terms
of \HI deficiency (Hoffman et al. 1988). Valluri \& Jog (1991) observed in
Virgo and some other rich clusters that galaxies with medium to large optical
sizes tend to be more severely \HI deficient compared to smaller galaxies in
terms of both the fractional number and the amount of gas lost. This behavior
is contrary to that expected from ram-pressure stripping or transport
processes, however, consistent with that expected if tidal interactions were
responsible for the gas deficiency.  The exact mechanism responsible for \HI
deficiency in clusters is still uncertain. These difficulties have led one to
speculate that cluster galaxies were perhaps \HI deficient even before they
fell into the cluster. Such a speculation is motivated by the hierarchical
theory of structure formation where clusters build up via mergers of small
groups. Groups of galaxies therefore provide an opportunity to trace early
evolution of galaxies.

Several clusters have been imaged in \HI. However, only limited \HI data exist
for large groups. Previous studies on groups were mainly aimed at Hickson
Compact Groups (HCG's), which usually have less than 10 galaxies packed in a
small volume . For instance, Verdes-Montenegro et al.  (2001) imaged several
HCG's in \HI and found a significant \HI deficiency in the galaxies. To our
knowledge, the only large group studied in \HI is Ursa-Major, which is  rich in
spirals, and has a few S0's and no ellipticals. Verheijen (2001) showed that
there is no \HI deficiency in the Ursa-Major galaxies. The environment in the
Ursa-Major group is similar to that in the field. The \HI data for groups
in which the environment is intermediate between field and cluster is 
lacking. Here,
we present an \HI study of the Eridanus group which appears to be an
intermediate system between a loose group like Ursa-major and a cluster like
Fornax or Virgo. The properties of the Eridanus group are described in detail
in  Omar \& Dwarakanath (2004; hereafter paper-I). The Eridanus group has a
significant population of S0's (paper-I). The origin of S0's has been the
subject of much debate. Usually the population of S0's is enhanced in clusters
where galaxy density is high. There are two hypotheses for the formation of
S0's, one is ``Nature'' where it is believed that S0's were formed as such, and
the other is ``Nurture'' (evolution) according to which these galaxies are
transformed spirals. The presence of enhanced population of S0's in the
Eridanus group indicates that significant evolution of galaxies has perhaps
already happened in the group. In the present study, \HI content of galaxies in
the Eridanus group is analysed. The aim is to identify the galaxy evolution
processes in the group environment. Both the GMRT data and the HIPASS (HI
Parkes All Sky Survey) data are used. The details of the GMRT observations,
data reduction and analyses are presented in paper-I and Omar (2004). Some new
results based on follow-up VLA observations are also presented here.

\section{The Eridanus group}

The Eridanus group was identified as a moderate size cluster in a large scale
filamentary structure near $cz\sim1500$~km~$s^{-1}$ in the Southern Sky
Redshift Survey (SSRS; da Costa et al. 1988). This filamentary structure, which
is the  most prominent in the southern sky, extends for more than 20 Mpc. The
Fornax cluster and the Dorado group of galaxies are also part of this
structure. Eridanus has  $\sim200$ galaxies distributed over $\sim10$~Mpc
region. The properties of the group are described in detail in paper-I. The
distance to the group is estimated as $\sim23\pm2$~Mpc. The group appears to be
made of different sub-groups which have different morphological mix. One of the
sub-group, NGC~1407 (cf. Willmer et al. 1989), has a population mix of (E+S0's) 
and (Sp+Irr's) in the ratios of 70\% \& 30\% respectively, which is quite similar to those
found in clusters. The overall population mix in the Eridanus group is 30\%
(E+S0) \& 70\% (Sp +Irr). These sub-groups often have their brightest member as
an elliptical or an S0. The brightest member in the entire group is a luminous
($L_{B} \sim 4 \times 10^{10}$ L$_{\odot}$) elliptical galaxy with diffuse
X-ray emission ($L_{\times;0.1-2.0 keV} \sim 2 \times 10^{41}$ erg s$^{-1}$)
surrounding it. Diffuse x-ray emission($L_{\times;0.1-2.0 keV} \sim 7 \times
10^{40}$ erg s$^{-1}$) is also seen around another elliptical galaxy NGC~1395
which belongs to another sub-group. There is no appreciable difference in the
velocities over which the early types and the late types are distributed. This
is contrary to that seen in Virgo, Coma and several nearby Abell clusters where
spirals have much flatter  velocity distribution while E/S0's have nearly a
Gaussian distribution  in velocity (Binggeli et al. 1987, Colless \& Dunn
 1996, Biviano et al. 2002). 

\begin{figure}
\centering
\includegraphics[width=8.5cm]{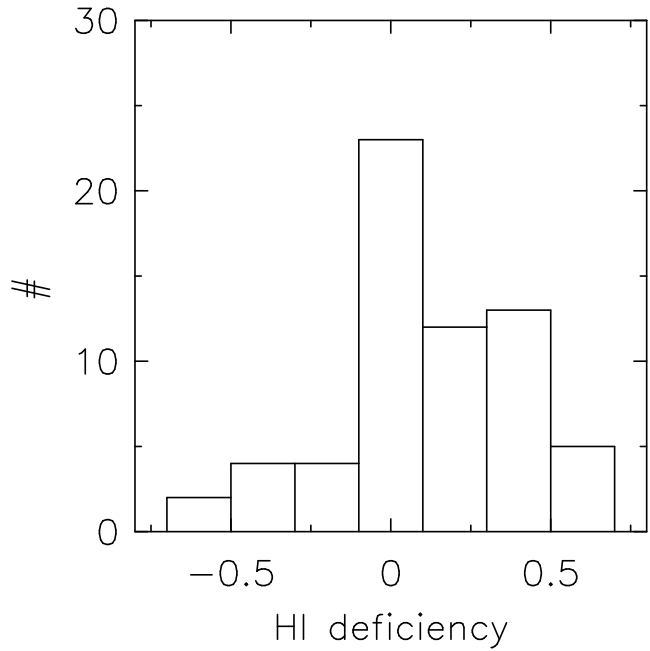}
\caption{Histogram of the \HI deficiency in the Eridanus galaxies.}
\label{fig:defi_histo}
\end{figure}

\begin{figure}
\centering
\includegraphics[width=10cm, angle=0]{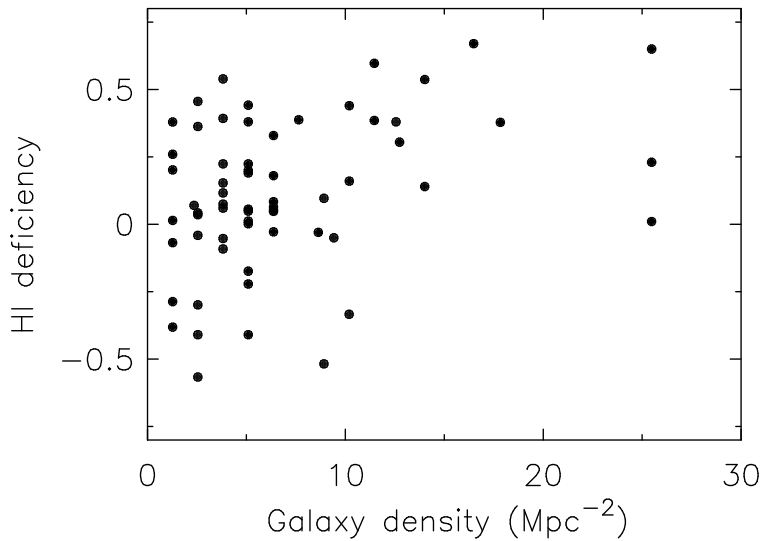}
\caption{\HI deficiency vs the local projected  galaxy density. The Spearman
Rank-Order Correlation Coefficient test shows that the correlation is significant at
$>99\%$.}
\label{fig:dens_defi}
\end{figure} 

\begin{figure}
\centering
\includegraphics[width=9cm, angle=0]{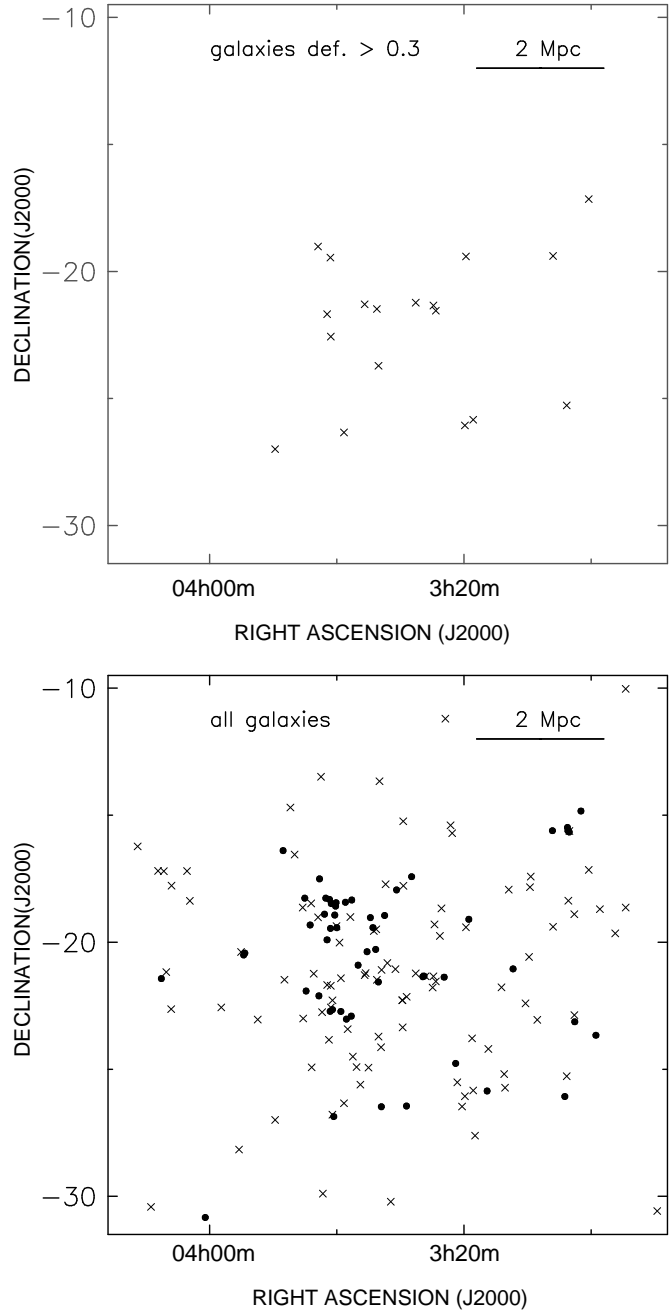}

\caption{(Lower panel) All galaxies in the Eridanus group. The early type
galaxies (E+S0) are marked as filled circles and late type galaxies (Sp+Irr)
are marked as crosses. (Upper panel) Galaxies with \HI deficiency greater than
0.3. It can be seen that severely \HI deficient galaxies and early type
galaxies are confined to the regions of higher galaxy densities.}

\label{fig:def_loc}
\end{figure}

\begin{figure}
\centering
\includegraphics[width=10cm, angle=0]{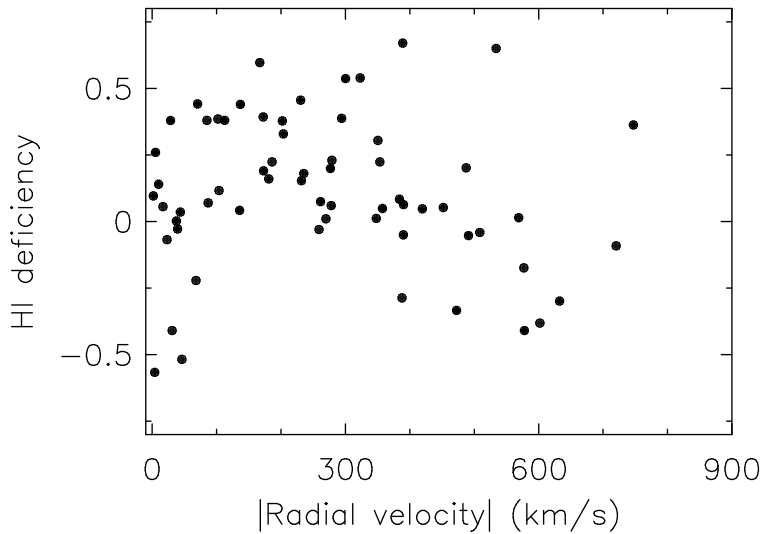}
\caption{\HI deficiency is plotted against the line-of-sight radial velocities
(w.r.t. the systemic velocity of the group) of galaxies in the group. The Spearman
Rank-Order Correlation Coefficient test shows that the correlation is significant at
$>99.9\%$ for velocities $>100$~\km.}
\label{fig:velo_defi}
\end{figure}

\section{The \HI content}

A total of 57 galaxies in the Eridanus group were observed with the GMRT in the
\HI 21 cm-line. The details of the observations and the data analyses are
described  in paper-I. The \HI detections were made for 31 galaxies. It was
noticed that the \HI flux densities of some large ($dia. > 6'$) galaxies were
underestimated by the GMRT observations presumably due to inadequate sampling
of the short $(u,v)$ spacings. In the present study, the \HI masses for such
galaxies were replaced by those obtained from the HIPASS data (Meyer et al.
2004). In addition, HIPASS data were used for galaxies in the Eridanus region
not observed by the GMRT. The final sample consisted of a total of 63 \HI detected
galaxies of different morphological types.  The \HI sample is described in
Appendix-A.

The \HI masses of galaxies in the field environment are observed to be
correlated with their Hubble types and optical diameters $D_{opt}$ (e.g.,
Haynes \& Giovanelli 1984, $hereafter$ HG84). The \HI deficiency (cf. HG84) for
a galaxy of a given type can be estimated by comparing $log
(M_{\HI}$/$D_{opt}^{2})$ of the galaxy with that observed for field galaxies.
In the present study, the ratio $log (M_{\HI}$/$D_{opt}^{2})$ for each Eridanus
galaxy is compared with the mean value of the ratio $log
(M_{\HI}$/$D_{opt}^{2})$ obtained by HG84 for isolated galaxies of similar
types. A significant positive difference between the two ratios indicates an
\HI deficiency (def. =  $<log(M_{\HI}/D_{opt}^{2})>_{field} -
log(M_{\HI}/D_{opt}^{2})$). It should be noted that this deficiency parameter
is distance independent. The optical diameters of
galaxies used in HG84 were from the the Upsala General Catalog (UGC). The
optical diameters of galaxies in the Eridanus group are from the Third
Reference Catalog of Galaxies (RC3; de Vaucouleurs et al. 1991). The optical
diameters in RC3 are at 25~mag~arc sec$^{-2}$ in the B-band. To convert the RC3
diameters or $D_{25}$ to $D_{opt}$ consistent with the UGC diameters, the
conversion relation obtained by Paturel et al. (1991) was used. This relation
predicts that the $D_{opt}$ (UGC) is about 1.09 times the $D_{25}$. 

A histogram of the \HI deficiency for the Eridanus galaxies is plotted in
Fig.~\ref{fig:defi_histo}. The \HI deficiency is independent of the
morphological type of the galaxies. It can be seen that although the
distribution peaks at zero deficiency, there are more galaxies with positive
differences. Some \HI rich galaxies (def. $< -0.5$) are also seen in
Fig.~\ref{fig:defi_histo}. These turn out to be interacting pairs, and hence
the \HI masses are likely to be overestimated.

Fig.~\ref{fig:dens_defi} shows the \HI deficiency plotted against the local
projected galaxy density. The projected galaxy density is estimated within a
circular region of diameter 1.0~Mpc. It can be seen that in  higher galaxy
density ($>10$~Mpc$^{-2}$) regions, majority of galaxies are \HI deficient
while in the lower galaxy density regions both normal and deficient galaxies
are present. Galaxies are \HI deficient up to a factor of $2-3$ ($\sim0.3-0.5$
in log units) in these plots.  The correlation between the projected galaxy
density and \HI deficiency can also be seen in Fig.~\ref{fig:def_loc} where the
locations of all identified group members (lower panel) and the locations of
galaxies deficient by more than a factor of two are plotted (top panel). It is
evident that severely \HI deficient galaxies and the early type galaxies are
confined to the regions of higher galaxy densities. The \HI deficiency also
shows a strong inverse correlation with the line-of-sight radial velocities
(w.r.t. the systemic velocity of the group) of
galaxies in the group (Fig.~\ref{fig:velo_defi}). All of these information are
used in the next section to identify the gas-removal mechanism active in the
group.

\section{The gas-removal processes}

Several gas-removal mechanisms have been discussed in the literature to explain
the \HI deficiency in cluster galaxies. These mechanisms are ram-pressure
stripping, transport processes (thermal conduction, and viscous and turbulent
stripping), galaxy harassment (tidal interactions), and galaxy strangulation
etc. All these mechanisms are expected to show some correlation of the \HI
deficiency with the properties of the cluster-environment and  of the galaxies.
For instance, ram-pressure stripping will be more effective for
galaxies with higher radial velocities in the group. The trend observed in
Fig.~\ref{fig:velo_defi} is opposite to that expected if ram-pressure stripping
were globally effective in the Eridanus group.  It can be shown that for
galaxies in the Eridanus group the ram-pressure is one to two orders of
magnitude lower than that in the cores of clusters (Omar 2004). This implies
that ram-pressure stripping is of a little importance in the Eridanus group.

The direct correlation of deficiency with the local projected galaxy density,
and the inverse correlation with the line-of-sight radial velocity suggest that the
\HI deficiency in Eridanus galaxies is due to tidal interactions. Galaxies in
higher galaxy density regions will have higher probability of tidal encounters.
Therefore, a direct correlation of deficiency with the local projected galaxy
density is expected. Further, the perturbation due to tidal interactions are
expected to be  maximum for slow encounters. In the Eridanus group where the
velocity distribution of galaxies is peaked near the mean velocity of the group
(paper-I) and falls off nearly as a  Gaussian at higher relative velocities,
galaxies having nearly zero radial velocities in the group will have a higher
probability of interacting with a companion having a lower velocity difference.
Therefore, the inverse correlation of \HI deficiency with the line-of-sight
radial velocity is qualitatively understood. The presence of a wider
distribution of deficiency near the zero radial velocity is likely to be due to
projection effects. Galaxies with higher radial velocities but moving nearly
perpendicular to the line-of-sight will have almost zero line-of-sight radial
velocities. Therefore, some discordant points are expected near zero velocity
in Fig.~\ref{fig:velo_defi}. 

\begin{figure}
\centering
\includegraphics[width=10cm, angle=0]{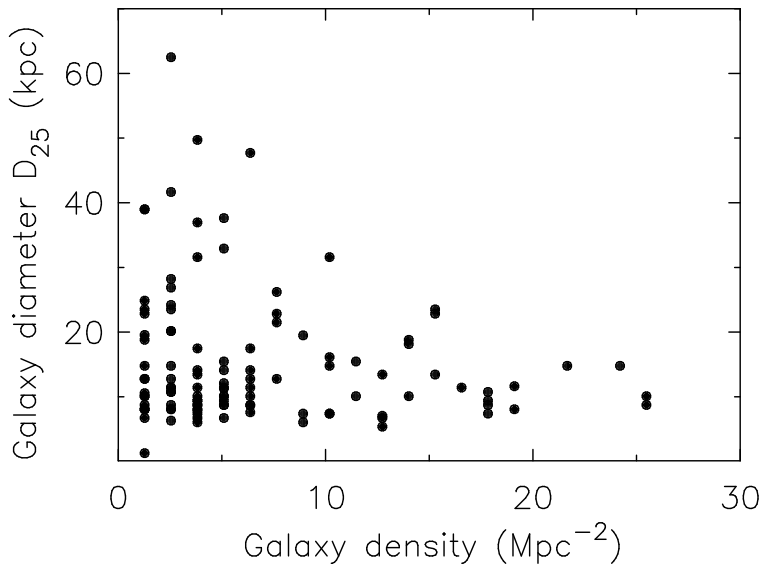}
\caption{The optical disk diameters plotted against the local projected galaxy density.}
\label{fig:size_dens}
\end{figure}

Tidal forces will affect both the gas and the stars in galaxies. 
In contrast, ram-pressure affects
only the gas. If the \HI deficiency in the Eridanus group is indeed due to tidal
interactions, some observational signatures of the same should be seen in both
the stellar and the \HI disks. The tidal interactions often produce gaseous and
stellar tidal tails extending to large distances in the IGM. Some of the gas
and the stars will be lost from the galaxy to the IGM in this process. It will be
difficult to detect this low column density tidal debris in the IGM except for
those associated with recent events where the column densities could still be
detectable. However, repeated tidal encounters in the high galaxy density
regions will shrink the optical sizes of galaxies. Fig.~\ref{fig:size_dens}
indicates that the Eridanus galaxies with larger optical sizes are
predominantly in the lower galaxy density regions. This trend is further
indicative of the scenario of tidal interactions being effective in the
Eridanus group. 

It is worthwhile to discuss an important effect while estimating \HI deficiency
using the $M_{\HI}/D^{2}_{opt}$ parameter. Since both $D_{opt}$ and $M_{\HI}$ are
reduced as a result of tidal encounters, the \HI gas loss inferred from this
deficiency parameter will be a lower limit. In the absence of detailed
simulations of repetitive tidal encounters in  a group environment, such
effects are hard to quantify.

\section{The morphological peculiarities of the Eridanus galaxies}

\begin{figure}
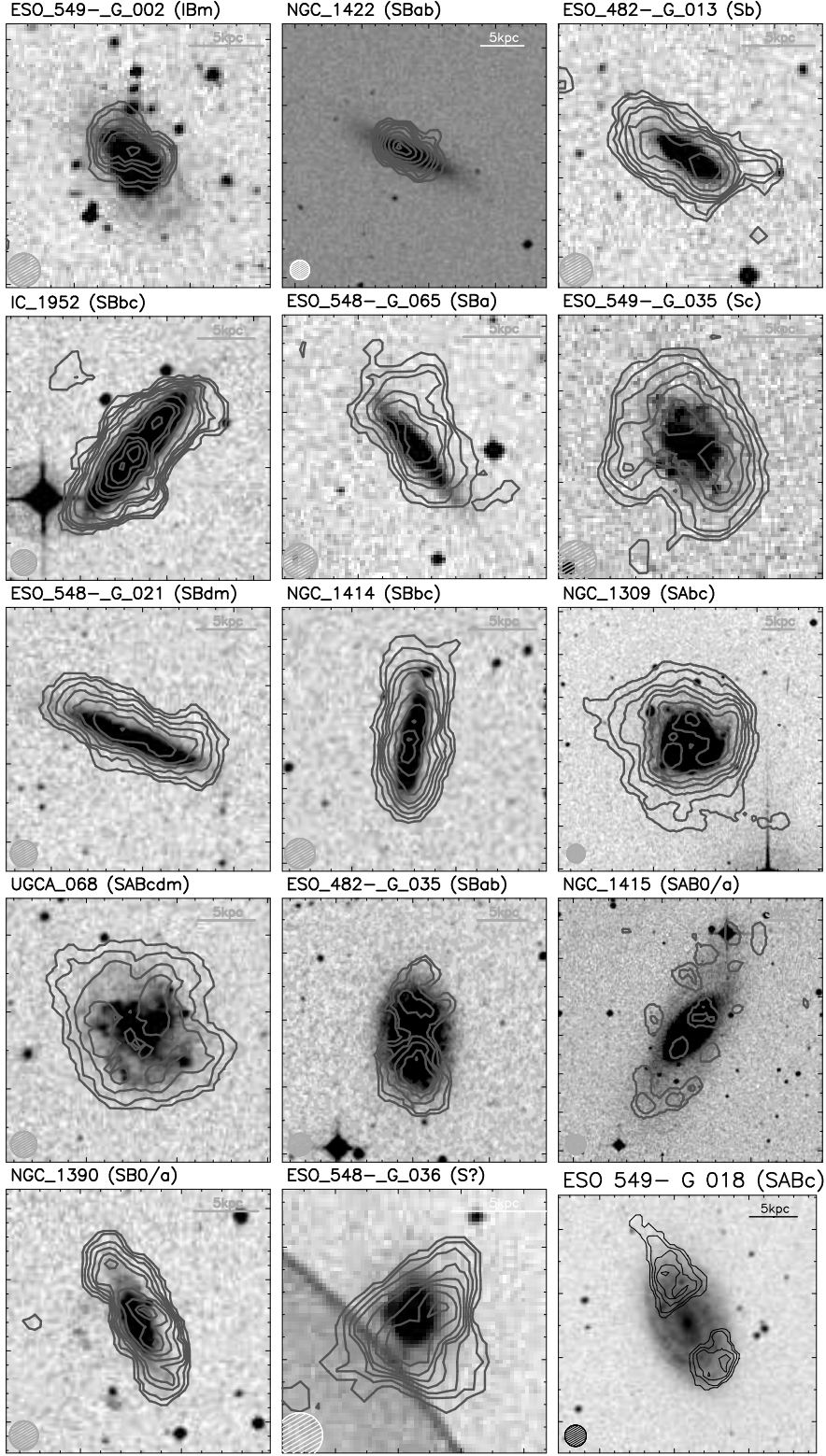

\centering

\includegraphics[width=3.8cm, angle=0]{ESO_549-_G_002.collage.epsi}
\includegraphics[width=3.8cm, angle=0]{NGC_1422.collage.epsi}
\includegraphics[width=3.8cm, angle=0]{ESO_482-_G_013.collage.epsi}
\includegraphics[width=3.8cm, angle=0]{IC_1952.collage.epsi}
\includegraphics[width=3.8cm, angle=0]{ESO_548-_G_065.collage.epsi}
\includegraphics[width=3.8cm, angle=0]{ESO_549-_G_035.collage.epsi}
\includegraphics[width=3.8cm, angle=0]{ESO_548-_G_021.collage.epsi}
\includegraphics[width=3.8cm, angle=0]{NGC_1414.collage.epsi}
\includegraphics[width=3.8cm, angle=0]{NGC_1309.collage.epsi}
\includegraphics[width=3.8cm, angle=0]{UGCA_068.collage.epsi}
\includegraphics[width=3.8cm, angle=0]{ESO_482-_G_035.collage.epsi}
\includegraphics[width=3.8cm, angle=0]{NGC_1415.collage.epsi}
\includegraphics[width=3.8cm, angle=0]{NGC_1390.collage.epsi}
\includegraphics[width=3.8cm, angle=0]{ESO_548-_G_036.collage.epsi}
\includegraphics[width=3.8cm, angle=0]{ESO_549-_G_018.collage.epsi}
\caption{GMRT \HI column density contours overlaid upon the optical DSS
 gray-scale images of Eridanus galaxies which show peculiar \HI morphologies
 (see Sect.~5 for details). The contours levels increase in units of $N_{\HI} =
2 X 10^{20}$~cm$^{-2}$. The first contour is at $N_{\HI} = 10^{20}$~cm$^{-2}$.}

\label{fig:collage}
\end{figure}

\begin{figure}
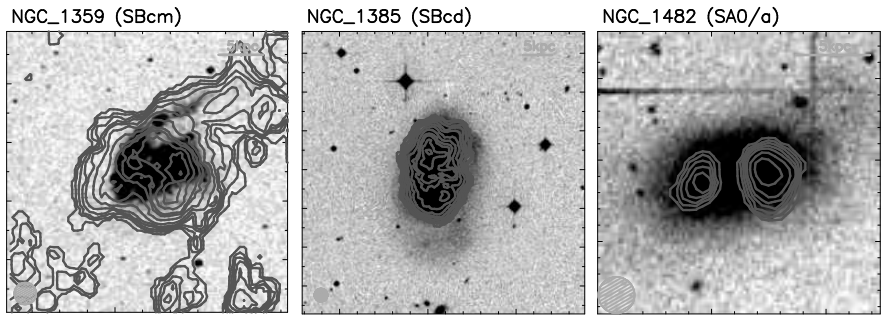

\centering
\includegraphics[width=3.8cm, angle=0]{NGC_1359.collage.epsi}
\includegraphics[width=3.8cm, angle=0]{NGC_1385.collage.epsi}
\includegraphics[width=3.8cm, angle=0]{NGC_1482.collage.epsi}

\caption{GMRT \HI column density contours overlaid upon the optical DSS
 gray-scale images of Eridanus galaxies which have clearly visible tidal
 tails in optical.}
\label{fig:collage2}
\end{figure}

Some galaxies in the Eridanus group show tidal tails and other peculiarities
like \HI extending out of the disk, \HI warps, asymmetric \HI disks, shrunken
or fragmented \HI disks, kinematical or  \HI lopsidedness etc. Tidal
interactions can produce long tails of gas and stars, and deformations in
the disks of galaxies. Some representative examples of these peculiarities are
shown in Figs.~\ref{fig:collage} \& \ref{fig:collage2}, and are discussed
below. 

\subsection{Shrunken \HI disks} 

ESO~549-~G~002  and NGC~1422 show shrunken \HI disks. Both of these galaxies
are \HI deficient, and are in a region with a galaxy density $\sim20$~Mpc$^{-2}$.
The \HI deficiencies for ESO~549-~G~002 and NGC~1422 are 0.67 and 0.44
respectively.  ESO~549-~G~002 has a faint stellar envelop in the outer region
and has an irregular optical morphology in the inner region. NGC~1422 is an
edge-on galaxy with a prominent dust lane. It has an \HI morphology similar to
that seen in ram-pressure stripped galaxies in clusters where gas from the
outer regions is preferentially removed. As it has been argued that the
ram-pressure is not much effective in the Eridanus group, it indicates that
highly shrunken \HI disk in NGC~1422 is due to some other reasons. Moreover, no
tidal features are seen either in optical or in \HI. The origin of shrunken \HI
disk in NGC~1422 therefore remains elusive. 

\subsection{Extra-planar gas \& warps} 

Some galaxies in the Eridanus group show \HI extending out of the disk (see ESO
482- G 013, IC 1952, ESO 548- G 065, ESO 549- G 035 in Fig.~\ref{fig:collage})
This extra-planar gas is seen in the form of wisps and small plumes of gas. It
appears that this phenomena is often seen in the edge-on galaxies since the
extra-planar gas will be easier to detect in these systems. It may be possible
that other galaxies also have such features. The extra-planar gas might be
quite common in galaxies. \HI warps can be noticed in ESO~548-G~021 and
NGC~1414. 

\subsection{Asymmetric \HI disks}

NGC~1309 and UGCA~068 have asymmetric \HI disks. These two galaxies have normal
\HI content. One side of the galaxy appears more diffuse than the opposite
side. These two galaxies also show kinematical asymmetries (Paper-I). NGC~1309
has a warp in the outer region. UGCA~068 shows asymmetry in the rotation curve.
We speculate that such features could be due to retrograde tidal encounters.
The retrograde encounter described by Toomre \& Toomre (1972) does not pull out
stars (the gas will respond in the same way) but can cause both morphological
and kinematical asymmetries.

\subsection{Peculiar \HI disks}

The \HI disks of ESO~482-G~035 and NGC~1415 are seen to be very peculiar. Both
are early type disk galaxies, and it is not common to see fully developed \HI
disks in such galaxies. It is likely that the  detected \HI is in a ring rather
than being in a fully developed disk. NGC~1415 is an S0/a galaxy with a faint
optical ring in the outer regions where most of the \HI is seen in isolated
clouds. The inclination of the ring is mis-aligned with the inner disk by more
than $20\deg$.   

The \HI disk of NGC~1390 is bent in an arc shape. Some diffuse nebulosity is
visible toward the west of this galaxy.

\subsection{Polar ring galaxy}

ESO~548~-G 036 (S?) has the position angle of the \HI disk inferred from \HI
kinematics almost normal to the position angle of the optical isophotes. The
optical body resembles an S0 galaxy with a dust lane normal to the major axis
of the main body.  It is suggested that this is a polar ring galaxy.  Polar
rings are believed to be due to recent accretion of gas from tidal encounters.
IC~1953, an \HI rich galaxy at  a projected separation of $\sim50$~kpc, is
likely to be the companion.

\subsection{Tidal tails}

Either gaseous or stellar tidal features are seen in  NGC~1359, NGC~1385, and
NGC~1482 (Fig.~\ref{fig:collage2}). The \HI tidal tail and some isolated \HI
features in the vicinity can be seen in NGC~1359. Both NGC~1385 and NGC~1482
are far-infrared luminous galaxies, and undergoing intense star forming
activities. NGC~1385 has highly asymmetric diffuse stellar envelop, however no
gaseous tidal tail is seen. NGC~1482 (S0/a) appears to have an \HI ring
coincident with the dust ring seen in the optical image. The apparent central
\HI hole in NGC~1482 is due to \HI absorption against the radio emission.
Stellar tidal features can be seen in NGC~1482. Follow up VLA \HI observations
on these galaxies were carried out aimed at detecting any low column density
tidal debris. The details are given in the next section.

\begin{table}
\begin{center}
\caption{VLA D-config. observations}
\vspace{0.1in}
\label{tab:vlaobs}
\begin{tabular}{llllcc}
\hline
\hline
\bf{\#} &\bf{Obs.} &\bf{Field centre} &\bf{Galaxies}  &\bf{rms}  &\bf{$\theta_{a} \times \theta_{b}$, PA} \\
       &\bf{date}       &($\alpha^{h,m,s}$, J2000)  &           &(mJy     &($^{''}~\times~^{''}, ~~\deg$) \\
       &(d-m-y)&($\delta^{d,',''}$, ~~J2000)  &           &/bm)     & \\
\hline
1   &30-05-04  &~03 ~54~ 49.0   &NGC 1482       &1.4 &$74\times57, 15.6$ \\
    &          &-20 ~26~ 14.0   &NGC 1481       & & \\
    &          &                              &ESO 549- G 035 & & \\
2   &05-06-04  &~03 ~40~ 31.5    &ESO 548- G 072 &1.0 &$67\times58, 04.5 $ \\
    &          &-19 ~25~ 00.0                   &ESO 548- G 065 & & \\
    &          &                              &ESO 548- G 064 & & \\
3   &11-06-04  &~03 ~41~ 12.5                   &NGC 1415       &1.0 &$76\times56, 15.1$ \\
    &          &-22 ~34~ 30.0                   &APM 482+009-132& & \\
    &          &                              &ESO 482- G 031 & & \\
    &          &                              &NGC 1416       & & \\
4   &12-06-04  &~03 ~33~ 34.8                   &ESO 548- G 036 &1.2 &$75\times56, 18.6$ \\  
    &          &-21 ~31~ 23.0                   &IC 1953        & & \\
5   &13-06-04  &~03 ~37~ 28.3                   &NGC~1385       &1.9 &$71\times57,$ -8.6 \\ 
    &          &-24 ~30~ 05.0                   &               &    &    \\
6   &13-06-04  &~03 ~34~ 00.0                   &NGC~1359       &1.9 &$84\times55, 27.0$ \\
    &          &-19 ~28~ 36.0                   &ESO 548- G 043 &    &    \\
    &          &                              &ESO 548- G 044 & & \\
\hline

\multicolumn{6}{p{5in}}{\ch Notes - (a) The bandwidth of each IF was 3.125~MHz. (b) Fields 1, 4, 5 and 6 were observed in 2 IFs  with 128 channels
in each IF. Fields 2 and 3 were observed in 4 IFs  with 64 channels in each IF.
The total integration time was $\sim3.2$ hrs for the fields 1, 2,
3, and 4 and $\sim1.6$ for fields 5 and 6.}

\end{tabular} 
\end{center} 
\end{table}

\begin{figure}
\centering
\includegraphics[width=10cm, angle=0]{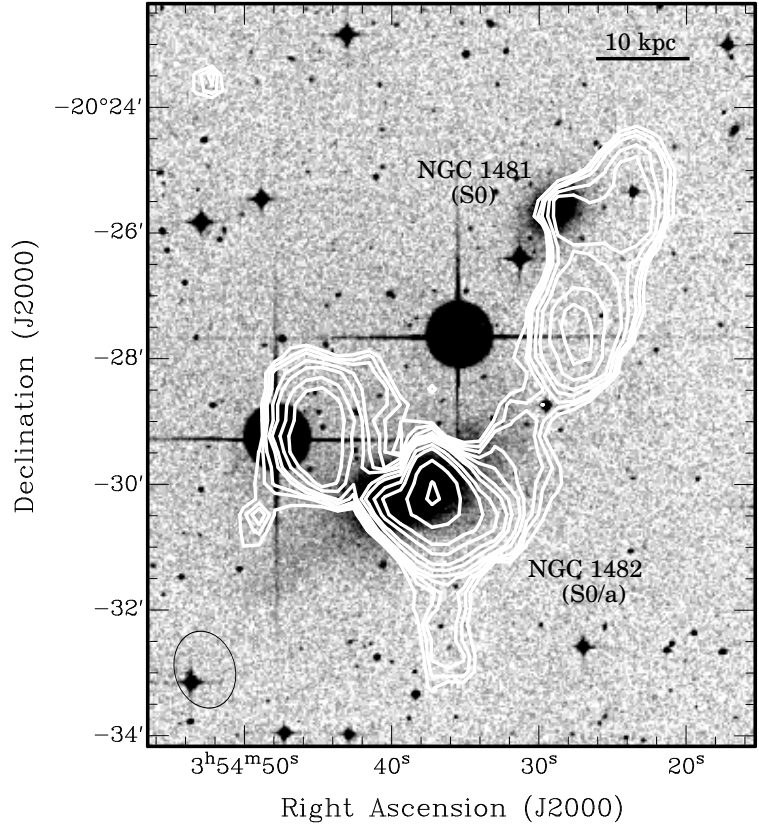}

\caption{Contours of the VLA \HI image of NGC~1482 overlaid upon the optical
image from DSS. The contours are at $N_{\HI}$ = 1, 1.5, 2, 3, 4, 6, 8, 12, 16,
24, 32 $\times10^{19}$cm$^{-2}$. Faint stellar streamers can be seen around NGC~1482.}

\label{fig:VLA1}
\end{figure}

\begin{figure}
\centering
\includegraphics[width=12cm, angle=0]{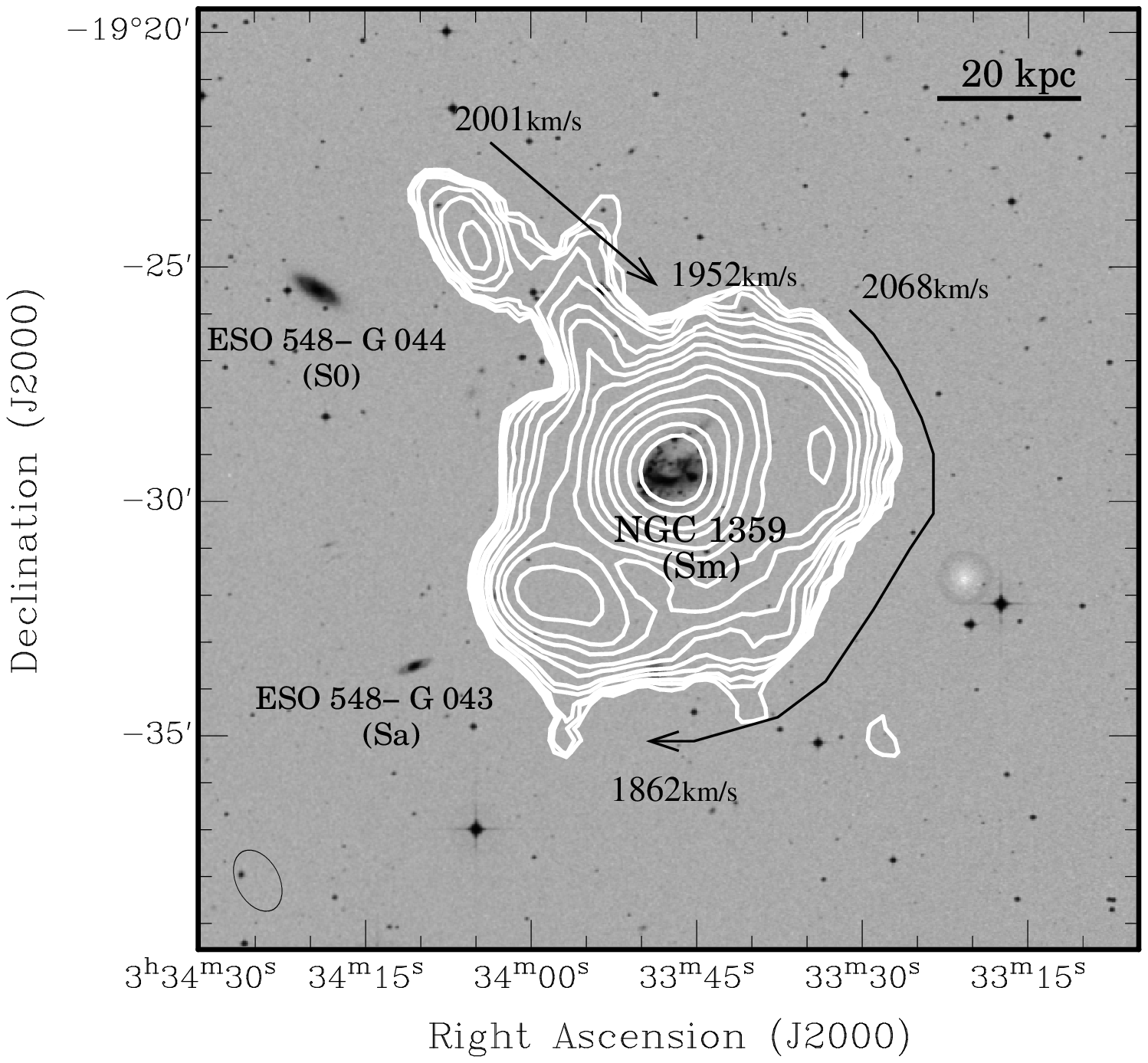}

\caption{Contours of the VLA \HI image of NGC~1359 overlaid upon the optical
image from DSS. The contours are at $N_{\HI}$ = 1, 1.5, 2, 3, 4, 6, 8, 12, 16,
24, 32, 48, 64, 96, 128 $\times10^{19}$cm$^{-2}$. Stellar tidal tail can be
seen in NGC~1359. The approximate directions of the velocity gradients in the
two tidal features are marked.}

\label{fig:VLA2}
\end{figure}

\section{Follow-up VLA observations}

Although several peculiarities are seen in both the optical and the GMRT \HI
images of Eridanus galaxies, no tidal debris in \HI were detected in the GMRT
images. It is expected that tidal debris will have low column density gas which
could have been missed in the GMRT images due to its limited sensitivity to the
extended emission. Follow-up VLA observations were, therefore, carried out in
its D-configuration which is most sensitive to the extended low column density
emission. The observations were carried out on galaxies having one or more
close neighbors or on galaxies which showed tidal features in optical.  The
details of the observations are given in Tab.~\ref{tab:vlaobs}. The
observations were carried out in the D-north-C (DnC) hybrid configuration which
gives a nearly symmetric synthesised beam for sources in the southern
hemisphere. The data were analysed following the standard procedures using {\sf
AIPS} (Astronomical Image Processing System) developed by the National Radio
Astronomy Observatory. The flux density scale is based on the standard VLA
calibrator 0137+331. Observations were carried out in two polarizations. Fields
2 and 3 (cf. Tab.~\ref{tab:vlaobs}) were observed in the 4IF correlator mode at
two different centre frequencies in the two IFs each of bandwidth 3.125~MHz
with sufficient overlap so that the total usable bandwidth after stitching the
two IFs was $\sim5$~MHz. Other fields were observed in the 2IF correlator mode
with 3.125~MHz of bandwidth. The velocity resolution was $\sim10.4$~\km for
fields 2 and 3, and $\sim5.2$~\km for the other fields. 

The \HI emission was searched by eye in the channel images. The \HI moment maps
were constructed using the {\sf AIPS} task {\sf MOMNT}. Field 1 (NGC~1482;
Fig.~\ref{fig:VLA1}) and field 6 (NGC~1359; Fig.~\ref{fig:VLA2}) show extended
\HI tails previously undetected in the GMRT images. The \HI images of other
galaxies are almost identical to those obtained from the GMRT. Both NGC~1482
and NGC~1359 are in the sub-groups of which they are the brightest members.
Both galaxies show tidal features in their optical images. It appears that
NGC~1482 has interacted with NGC~1481. NGC~1359 has two nearest neighbors, but
\HI streamers do not connect them. It appears that NGC~1359 has undergone
multiple interactions as there are two different \HI streamers or tails, one
toward the north-east and the other toward the west winding toward the south.
An understanding of the \HI morphologies in these galaxies requires detailed
N-body simulations.  Nevertheless,  these \HI detections indicate that tidal
interactions are effective in the Eridanus group.

\section{Discussion}

Galaxies in the Eridanus group are \HI deficient up to a factor of $2-3$
compared to their field counterparts. The direct correlation of the \HI
deficiency with the local galaxy density, and the inverse correlation of the
\HI deficiency with the line-of-sight radial velocities of galaxies suggest
that the deficiency is due to tidal interactions. Since the \HI deficiency of
the Eridanus galaxies is observed in all types of galaxies, the correlation of
the \HI deficiency with the local galaxy density can not be just a
manifestation of the density-morphology relation. Although Eridanus appears as
a loose group,  it has significant sub-grouping. These sub-groups have regions
of higher galaxy densities where \HI deficient galaxies are seen. A Galaxy
moving with a radial velocity of $\sim240$ km s$^{-1}$ (velocity dispersion of
the galaxies in the group) can cross a linear distance of $\sim1$~Mpc (typical
extents of the sub-groups) in $\sim5$~Gyr (typical ages of the galaxies). The
higher galaxy density and the relatively short crossing time in the sub-groups
enhance the chances of encounters between the galaxies. Therefore, it appears
that the sub-grouping is playing an important role in producing the \HI
deficiency. Further, the inverse correlation of deficiency with the radial
velocity indicates that it is necessary for a galaxy to interact with another
galaxy with smaller velocity difference to make the tidal encounters
effective. 

The observed \HI deficiency in a loose group like Eridanus has some interesting
implications for the evolution of galaxies. Galaxies in clusters are known to
be \HI deficient up to a factor of 10. However, the reasons for such large
deficiencies are not completely understood. If clusters form as mergers of groups
as expected in a hierarchical Universe, a fraction of the \HI deficiency seen
in cluster galaxies might have originated in the group environment. In other
words, clusters are built with galaxies, some of them are already \HI
deficient. This implication is particularly important for simulations which try
to match the observed \HI deficiency in cluster galaxies with what can be
produced by ram-pressure stripping in the cluster environment (e.g., Vollmer et
al. 2001). Tidal interactions will become less effective in removing matter
from galaxies in a cluster environment as encounters will be  relatively fast.
However, increased frequency of encounters in clusters due to
higher galaxy density may actually affect the outer regions of galaxies as
envisaged by the process of ``galaxy harassment''. Therefore, it is possible
that tidal interactions play an important role in the evolution of galaxies in both
the group and the cluster environments. 

The origin of S0's in high galaxy density regions is not understood. Whether
S0's are due to some evolutionary processes driven by the environment
(``$Nurture$'') or due to the result of natural processes of galaxy formation
(``$Nature$'') is being debated. Several co-workers (Poggianti et al. 1999,
Dressler et al. 1997, Fasano et al. 2000) found  in observations of clusters at
intermediate redshifts ($z \sim 0.1-0.3$) that the fraction of S0's  tends to
grow at the expense of the spiral population as the redshift decreases. These
observations support the ``$Nurture$'' scenario. Several mechanisms have been
proposed for the transformation of spirals into S0's, e.g.,  ram-pressure
stripping, galaxy harassment, strangulation. The present observations in the
Eridanus group indicate that both the severely \HI deficient galaxies and the S0's
are found in the higher galaxy density regions in the Eridanus group. It may be
an indication that both the \HI deficient and the S0 galaxies originated through
similar  processes. Since the ram-pressure is not playing any major role in the
Eridanus group, the S0's in Eridanus are produced by some other process.
Tidal interactions appear to be a favorable mechanism for transformation of
spirals into S0's in the wake of the current understanding of the Eridanus group
from this study. 

The \HI deficiency observed in the Eridanus galaxies is consistent with the \HI
deficiency seen in other low velocity dispersion groups and clusters,
viz., the Hickson Compact groups (Verdes-Montenegro et al. 2001) and the Fornax 
cluster (Schroder et al. 2001). Mulchaey (2000) observed that groups of
galaxies are filled with the metal-enriched intra-group medium with average
metallicity $\sim0.3$.  While other mechanisms can also enrich the intra-group
medium with metals, gas lost from the galaxies via tidal interactions will have
some contribution to the metal content of the intra-group medium.

The environment in the Eridanus group is intermediate between that in a loose
group like the Ursa-Major and a cluster like the Fornax or the Virgo (paper-I). The
Ursa-Major has a few S0's and the \HI contents of spirals are similar to the
field spirals. However, the Eridanus group which has relatively large velocity
dispersion, relatively large fraction of S0's, shows significant \HI
deficiency. Willmer et al. (1989) indicated that the Eridanus group is in its early phase
of  cluster formation. If all these results are combined, it appears that the
Eridanus group is indeed in an initial phase of cluster
formation where physical conditions in the group are favorable for driving
the galaxy evolution.

\section{Conclusions}

\begin{itemize}

\item The galaxies in the Eridanus group are \HI deficient up to a factor of $2-3$.
The \HI deficiency is inversely correlated with the line-of-sight radial
velocity, and directly correlated with the local projected galaxy density.

\item The \HI deficiency in the Eridanus group is likely to be due to tidal
interactions.

\item The optical diameters of galaxies are observed to be reduced in the high
galaxy density regions indicating the effectiveness of tidal interactions in
the Eridanus group.

\item If clusters are built via mergers of groups, a fraction of the \HI deficiency
in cluster galaxies might have been produced in the group environment. 

\item The co-existence of S0's and \HI deficient galaxies in the Eridanus group
suggests that tidal interactions may be an effective mechanism for transforming
spirals to S0's.

\item The environment in the Eridanus group is intermediate between the
field and a cluster. Nevertheless, conditions are favorable for driving
galaxy evolution in the Eridanus group.

\end{itemize}

\section*{Acknowledgments} 

We thank Marc Verheijen, Jacqueline van Gorkom, and the referee for useful comments. This research has made use of the
HI Parkes All Sky Survey (HIPASS) data. This research has been benefited by
the NASA's Astrophysics Data System (ADS) and Extra-galactic Database (NED)
services.

\appendix
\section{Table : The \HI detected galaxies}

\noindent{\it Column~1} -- {\sf Galaxy name} \\
\noindent{\it Column~2} -- {\sf Hubble Type} \\
\noindent{\it Column~3} -- {\sf UGC optical diameter} \\
\noindent{\it Column~4} -- {\sf Systemic velocity} \\
\noindent{\it Column~5} -- {\sf Projected galaxy density} \\
\noindent{\it Column~5} -- {\sf \HI mass} \\
\noindent{\it Column~5} -- {\sf \HI deficiency} \\

\begin{table}
\begin{center}
\caption{The \HI detected galaxies}
\label{tab:data}
\begin{tabular}{lcccccl}
\hline
\hline
\ch {\bf Galaxy}& \ch {\bf H.T.}& \ch {\bf D$_{UGC}$}& \ch {\bf Velocity}& \ch
{\bf Proj. density} & \ch {\bf log(M$_{\HI}$/M$_{\odot}$)}& \ch{\bf Deficiency} \\
 &  & \ch (kpc) & \ch (\km) & \ch (Mpc$^{-2}$) & &  \\
\hline
NGC~1076& 	2& 13.9& 2102&  1.3& 9.27& -0.38 \\ 
NGC~1069& 	3& 25.5& 1456&  2.5& 9.47&  0.04 \\ 
NGC~1140& 	9& 12.4& 1496&  2.5& 9.62& -0.57 \\ 
ESO~546-~G~034& 9&  9.5& 1568&  5.1& 9.05& -0.22 \\ 
NGC~1163& 	6& 21.2& 2247&  2.5& 9.14&  0.36 \\ 
NGC~1187& 	7& 40.1& 1396&  3.8& 9.88&  0.12 \\ 
NGC~1179& 	8& 35.7& 1777&  5.1& 9.78&  0.20 \\ 
UGCA~050& 	9& 13.9& 1731&  2.5& 8.70&  0.46 \\ 
UGCA~051& 	9& 12.4& 1672&  3.8& 8.66&  0.39 \\ 
ESO~547-~G~011& 9&  9.5& 2220&  3.8& 8.92& -0.09 \\ 
IC~1898& 	7& 26.3& 1327&  5.1& 9.44&  0.19 \\ 
ESO~547-~G~020& 9& 10.2& 1991&  3.8& 8.94& -0.05 \\ 
NGC~1255& 	6& 30.6& 1686&  3.8& 9.60&  0.22 \\ 
UGCA~061& 	9& 29.2& 1735&  6.4& 9.62&  0.18 \\ 
NGC~1258& 	8&  9.5& 1483&  5.1& 8.77&  0.06 \\ 
UGCA~063& 	9&  8.8& 2077&  5.1& 8.93& -0.17 \\ 
NGC~1292& 	7& 21.9& 1364&  2.5& 9.43&  0.04 \\ 
UGCA~064& 	8& 21.1& 1794&  7.6& 9.13&  0.39 \\ 
UGCA~065& 	9& 14.6& 1537&  5.1& 9.20&  0.00 \\ 
NGC~1300& 	6& 45.2& 1571&  5.1& 9.72&  0.44 \\ 
NGC~1302& 	3& 28.4& 1704&  6.4& 9.27&  0.33 \\ 
NGC~1306& 	5&  8.0& 1454&  8.9& 9.16& -0.52 \\ 
MCG~-03-09-027&	9& 11.7& 2009&  2.5& 9.05& -0.04 \\ 
NGC~1309& 	6& 16.0& 2132&  2.5& 9.56& -0.30 \\ 
UGCA~068& 	9& 12.4& 1848&  5.1& 9.05&  0.01 \\ 
NGC~1325& 	6& 34.3& 1602& 11.5& 9.53&  0.39 \\ 
NGC~1325A& 	6& 16.0& 1333& 11.5& 8.66&  0.60 \\ 
ESO~548-~G~021& 8& 14.6& 1702& 17.8& 8.82&  0.38 \\ 
NGC~1345& 	7& 10.9& 1531&  5.1& 9.28& -0.41 \\ 
UGCA~075& 	9& 13.9& 1888&  1.3& 9.44& -0.29 \\ 
UGCA~077& 	9& 12.4& 1952&  6.4& 9.00&  0.05 \\ 
ESO~482-~G~005& 8& 12.3& 1920&  6.4& 9.00&  0.05 \\ 
NGC~1357& 	4& 20.4& 1987&  1.3& 9.10&  0.20 \\ 
IC~1952& 	6& 19.0& 1823&  3.8& 8.87&  0.54 \\ 
IC~1953& 	8& 20.4& 1851& 12.7& 9.19&  0.30 \\ 
NGC~1359& 	9& 17.5& 1973& 10.2& 9.69& -0.33 \\ 
NGC~1371& 	3& 40.8& 1461&  6.4& 9.94& -0.03 \\ 
IC~1962& 	8& 19.7& 1800& 14.0& 8.92&  0.54 \\ 
ESO~482-~G~013& 5&  8.2& 1854&  5.1& 8.44&  0.22 \\ 

\end{tabular} 
\end{center} 
\end{table}

\begin{table}[h]
\begin{center}
\begin{tabular}{lcccccl}
\hline
\ch {\bf Galaxy}& \ch {\bf H.T.}& \ch {\bf D$_{UGC}$}& \ch {\bf Velocity}& \ch
{\bf Proj. density} & \ch {\bf log(M$_{\HI}$/M$_{\odot}$)}& \ch{\bf Deficiency} \\
 &  & \ch (kpc) & \ch (\km) & \ch (Mpc$^{-2}$) & &  \\
\hline
NGC~1385& 	8& 24.8& 1498&  8.9& 9.56&  0.10 \\ 
NGC~1390& 	3& 10.2& 1230& 25.5& 8.70&  0.01 \\ 
NGC~1398& 	4& 51.8& 1388&  5.1& 9.74&  0.38 \\ 
ESO~482-~G~032& 9& 10.9& 1732&  3.8& 8.79&  0.15 \\ 
NGC~1425& 	5& 42.3& 1505&  1.3& 9.82&  0.26 \\ 
NGC~1421& 	6& 25.5& 2069&  1.3& 9.65&  0.01 \\ 
MCG~-03-10-045& 9&  9.5& 1239&  3.8& 8.75&  0.07 \\ 
UGCA~085& 	7& 24.8& 1529&  1.3& 9.20&  0.38 \\ 
ESO~549-~G~035& 7& 10.2& 1794&  3.8& 9.05& -0.24 \\ 
IC~2007& 	7&  9.5& 1523&  1.3& 8.81& -0.07 \\ 
SGC~0401.3-1720& 9& 12.4& 1890&  6.4& 8.99&  0.06 \\ 
UGCA~087& 	9& 16.8& 1884&  6.4& 9.24&  0.08 \\ 
NGC~1518& 	8& 21.9&  922&  2.5& 9.96& -0.41 \\ 
UGCA~088& 	8& 13.9& 1858&  5.1& 9.10&  0.05 \\ 
ESO~548-~G~049&  ?&  6.7& 1510& 14.0& 8.47&  0.14 \\
ESO~549-~G~035&  5&  9.4& 1778&  3.8& 8.72&  0.06 \\
ESO~548-~G~065&  1& 10.1& 1221& 25.5& 8.46&  0.23 \\
NGC~1347&        5&  7.1& 1759&  8.6& 8.64& -0.03 \\
NGC~1415&        1& 25.8& 1585& 12.6& 9.05&  0.38 \\
NGC~1414&        4& 11.6& 1681& 10.2& 8.76&  0.16 \\
ESO~548-~G~072&  5&  8.7& 2034& 25.5& 8.30&  0.65 \\
ESO~482-~G~035&  2& 12.8& 1890&  9.4& 8.64& -0.05 \\ 
NGC~1422&        3& 14.8& 1637& 10.2& 8.49&  0.44 \\
ESO~549-~G~002&  9&  8.7& 1111& 16.5& 8.27&  0.67 \\
ESO~549-~G~018&  5& 17.5& 1587&  2.4& 8.63&  0.07 \\
\hline
\end{tabular} 
\end{center} 
\end{table}


\begin{thebibliography}{}
\bibitem{bing} Binggeli, B.,  Tammann, G.A., \& Sandage, A. 1987, {\it \AJ}, {\bf 94}, 251
\bibitem{ESO} Biviano, A., Katgert, P., Thomas, T., \& Adami, C. 2002, {\it \aa}, {\bf 387}, 8
\bibitem{bravo} Bravo-Alfaro, H., Cayatte, V., van Gorkom, J. H., \& Balkowski, C. 2000, {\it \AJ}, {\bf 119}, 580
\bibitem{cay} Cayatte, V., van Gorkom, J. H., Balkowski, C., \& Kotanyi, C. 1990, {\it \AJ}, {\bf 100}, 604
\bibitem{col} Colless, M. \& Dunn, A. M. 1996, {\it \ApJ}, {\bf 458}, 435
\bibitem{Cowie} Cowie, L. L., \& Songaila, A. 1977, {\it Nature}, {\bf 266}, 501
\bibitem{daCos88} da Costa, L.N., Pellegrini, P.S., Sargent, W.L. et al. 1988, {\it \ApJ}, {\bf 327}, 544
\bibitem{dav} Davies, R. D. \& Lewis, B. M. 1973, {\it \MNRAS}, {\bf 165}, 231
\bibitem{devac} de vaucouleurs et al. 1991, Third Reference Catalogue of Bright
Galaxies, Springer Verlag, vol. 1-3
\bibitem{dres2} Dressler, A., Oemler, A. Jr., Couch, W. J. et al. 1997, {\it \ApJ}, {\bf 490} 577
\bibitem{fasano} Fasano, G., Poggianti, B. M., Couch, W. J. et al. 2000, {\it \ApJ}, {\bf 542}, 673
\bibitem{GH85} Giovanelli, R., \& Haynes, M. P. 1985, {\it \ApJ}, {\bf 292}, 404
\bibitem{gun} Gunn, J.E., \& Gott, J.R. 1972, {\it \ApJ}, {\bf 176}, 1
\bibitem{HG84} Haynes, M. P., \& Giovanelli, R.  1984, {\it \AJ}, {\bf 89}, 758
\bibitem{hoffman} Hoffman, G. L., Helou, G., \& Salpeter, E. E. 1988, {\it \ApJ}, {\bf 324}, 75
\bibitem{larson} Larson R.B. Tinsley, B.M., \& Calswell, C.N. 1980,{\it \ApJ}, {\bf 237}, 692
\bibitem{magri} Magri, C., Haynes, M. P., Forman, W., Jones, C., \& Giovanelli, R. 1988,
{\it \ApJ}, {\bf 333}, 136
\bibitem{hipass} Meyer, M.J., Zwaan, M.A., Webster, R.L. et al. 2004, {\it \MNRAS}, {\bf 350}, 1195
\bibitem{moore} Moore, B., Lake, G., \& Katz, N. 1998, {\it \ApJ}, {\bf 495}, 139
\bibitem{mul00} Mulchaey, J. S. 2000, {\it \ARAA}, {\bf 38}, 289
\bibitem{Nulsen} Nulsen, P.E.J. 1982, {\it \MNRAS}, {\bf 198}, 1007
\bibitem{omar} Omar, A. 2004, {\it Ph.D. thesis, Jawaharlal Nehru University, Delhi}
\bibitem{omar} Omar A. \& Dwarakanth K. S. 2004, {\it J. Astroph. Astron.},
this volume, paper-I
\bibitem{pati} Paturel, G., Garcia, A.M., Fouque, P., \& Buta, R. 1991, {\it \aa}, {\bf 243}, 319
\bibitem{pog} Poggianti, B. M., Smail, I., Dressler, A. et al. 1999, {\it \ApJ}, {\bf 518}, 576
\bibitem{sar} Sarazin C.L. 1988, X-ray emission from clusters of galaxies, Cambridge
Astrph. Series, Cambridge University Press.
\bibitem{sch} Schroder, A., Drinkwater, M. J., \& Richter, O.-G. 2001, {\it \aa}, {\bf 376}, 98
\bibitem{sol01} Solanes, J. M., Manrique, A., García-Gómez, C. et al.  2001, {\it \ApJ}, {\bf 548}, 97
\bibitem{toomre} Toomre, A., \& Toomre, J. 1972, {\it \ApJ}, {\bf 178}, 623
\bibitem{valuuri} Valluri, M. \&  Jog, C. J. 1991, {\it \ApJ}, {\bf 374}, 103
\bibitem{vanG03} van Gorkom, J.H. 2003, {\it In clusters of galaxies: Probes of cosmological
structure and galaxy formation, ed. Mulchaey, J.S, Dressler, A., \& Oemler, A., (Carnegie Obs. Astroph. Ser.)}, {\bf 3}
\bibitem{HCG} Verdes-Montenegro, L., Yun, M. S., Williams, B. A., Huchtmeier, W. K., Del
Olmo, A., \& Perea, J. 2001, {\it \aa}, {\bf 377}, 812
\bibitem{ver} Verheijen, M.A.W, 2001, {\em In gas and galaxy evolution, ed. Hibbard, J.E. et al., ASP
conf. series}, {\bf 240}, 573
\bibitem{vol01} Vollmer, B., Cayatte, V., Balkowski, C., \& Duschl, W. J. 2001, {\it \ApJ},
{\bf 561}, 708 
\bibitem{wil89} Willmer, C. N. A., Focardi, P., Da Costa, L. N., \& Pellegrini, P. S. 1989,
{\it \AJ}, {\bf 98}, 1531


\end{thebibliography}
\end{document}